\documentclass[12pt,preprint2]{emulateapj}
\usepackage{pslatex}
\usepackage[T1]{fontenc}
\usepackage[latin1]{inputenc}
\usepackage{subfigure}
\usepackage{amsmath}
\usepackage{graphicx,graphics}
\usepackage{amssymb}
\usepackage[english]{babel}

\newcommand{\be}{\begin{equation}}
\newcommand{\ee}{\end{equation}}
\newcommand{\beqn}{\begin{eqnarray}}
\newcommand{\eeqn}{\end{eqnarray}}
\def\gap{\;\rlap{\lower 2.5pt
 \hbox{$\sim$}}\raise 1.5pt\hbox{$>$}\;}
\def\lap{\;\rlap{\lower 2.5pt
   \hbox{$\sim$}}\raise 1.5pt\hbox{$<$}\;}
\def\mh{M_{BH}}

\shorttitle{Cusps around Black Holes }
\shortauthors{Preto, Merritt, Spurzem}
\begin{document}

\title{$N$-Body Growth of a Bahcall-Wolf Cusp around a Black Hole}

\author{Miguel Preto}
\affil{Department of Physics and Astronomy, Rutgers University, New Brunswick, 
NJ 08903, USA}

\author{David Merritt}
\affil{Department of Physics, Rochester Institute of Technology, 
Rochester, NY 14623, USA}

\and

\author{Rainer Spurzem}
\affil{Astronomisches Rechen-Institut Heidelberg, Mönchhofstrasse 12--14, 69120
Heidelberg, Germany}

%\date{\today}

\begin{abstract}

We present a clear $N$-body realization of the growth of
a Bahcall-Wolf $f\propto E^{1/4}$ ($\rho\propto r^{-7/4}$) 
density cusp around a massive object (``black hole'') at the 
center of a stellar system.
Our $N$-body algorithm incorporates a novel implementation of
Mikkola-Aarseth chain regularization to handle close
interactions between star and black hole particles.
Forces outside the chain were integrated on a GRAPE-6A/8 
special-purpose computer with particle numbers up to  
$N=0.25\times 10^6$.
We compare our $N$-body results with predictions of
the isotropic Fokker-Planck equation and verify that
the time dependence of the density (both configuration
and phase-space) predicted by the Fokker-Planck equation
is well reproduced by the $N$-body algorithm, 
for various choices of $N$ and of the black hole mass.
Our results demonstrate the feasibility of direct-force integration
techniques for simulating the evolution of galactic nuclei on 
relaxation time scales.

\end{abstract}

\keywords{stellar dynamics, galaxies: nuclei, black holes}

\section{Introduction}

The distribution of stars around a massive black hole is
a classic problem in galactic dynamics.
Many such distributions are possible, depending on the initial 
state,
the mode of formation of the black hole, and the time
since its formation.
However, a number of plausible scenarios predict a steeply-rising
stellar density within the black hole's sphere of influence,
$r\lap r_h\equiv G\mh/\sigma^2$, with $\mh$ the black hole
mass and $\sigma^2$ the 1D stellar velocity dispersion
outside of $r_h$.
Such models receive support from the observed run of stellar
density with radius near the centers of the nearest galaxies
known to contain massive black holes: the Milky Way, M31 and
M32.
The nucleus of each galaxy has a density cusp with 
$\rho\sim r^{-\gamma}$, $\gamma\approx 1.5$ within
the black hole's sphere of influence \citep{Lauer:98,Genzel:03}.
This  is consistent with the  slope predicted by  so-called 
adiabatic growth models, in which a black  hole with small
initial mass is embedded in a star cluster and its mass increased
to some final value, on a time scale long compared with 
orbital periods.
The final density in the adiabatic growth model
follows a power law at $r\lap r_h$, with index $\gamma_f$ that
depends on the initial stellar distribution function.
If the initial model is a non-singular isothermal sphere,
then $\gamma_f=3/2$ \citep{Peebles:72a,young80}.
However a non-singular isothermal sphere seems a rather ad hoc
guess for the initial state, and initial models without
flat cores give steeper final slopes, 
$1.5\lap\gamma_f\lap 2.5$ \citep{Merritt:04}.

Uncertainties in the initial state are less consequential
if the stellar cluster is old compared with the time $T_r$ for 
gravitational scattering to redistribute energy between stars.
In this case, one expects the collisional transport of mass
and energy to set up a steady-state distribution whose functional
form is
indendent of the initial phase-space density.
Peebles (1972b) first addressed this problem and derived
a power-law index $\gamma=9/4$ for the stellar density within
$r_h$.
Peebles obtained his solution by setting the flux of
stars in energy space to a constant non-zero value.
Shapiro \& Lightman (1976) and Bahcall \& Wolf (1976)
criticized the Peebles derivation on the grounds that the
implied flux is unphysically large, in fact divergent
if the solution is extended all the way to the black hole.
A physically more reasonable solution would have a
nearly zero  flux of stars into the black hole.
Bahcall \& Wolf (1976) repeated Peebles's derivation,
setting the phase-space density $f(E)$ to zero at the
black hole's tidal disruption radius.
They found that $f$ evolves, in a time only slightly longer
than $T_r$, to a steady state in which the flux is close
to  zero at all energies.
The zero-flux solution has $f\propto E^{1/4}$ within
the black hole's sphere of influence, with $E\ge 0$
the binding energy; the corresponding stellar density is
$\rho\propto r^{-7/4}$.

The Bahcall-Wolf solution has been verified in a number
of subsequent studies, almost all of which were
based on  the Fokker-Planck formalism
\citep{cohnk78,marchants80,freitagb02,amaroetal04}.
The $f\propto E^{1/4}$ and $\rho\propto r^{-7/4}$
character of the solution has been found to be robust, at
least at radii where the capture
or destruction  of stars occurs in a time long
compared with orbital periods.
But verifying the Bahcall-Wolf solution in an $N$-body integration
is also clearly desirable,
since an $N$-body simulation is free of many of the
simplifying assumptions that limit the applicability
of  the Fokker-Planck equation, including the restriction
to small-angle scattering, and the neglect of spatial
inhomogeneities in the derivation of the diffusion coefficients.
But the $N$-body approach is challenging:  large particle
numbers are required to resolve  the cusp, and accurate
integration schemes are needed to follow the
motion of stars near the black  hole.

In this paper we combine a sophisticated $N$-body code
with the special-purpose GRAPE hardware and show that
the formation of a Bahcall-Wolf cusp can be convincingly
reproduced without any of the approximations that
go  into the Fokker-Planck formalism.
Our results provide a clear demonstration of  the 
applicability of direct-force
$N$-body techniques to the  collisional evolution  of
dense star clusters,
and highlight the usefulness of direct $N$-body techniques 
for understanding the dynamical evolution of
galactic nuclei containing supermassive black holes.

\section{Method}

Our $N$-body algorithm is an adaptation of NBODY1 \citep{aarseth99a}
to the GRAPE-6 special-purpose hardware.
Forces for the majority of the particles were computed via
a direct-summation scheme without regularization.
Particle positions were advanced using a fourth-order
Hermite integration scheme with individual, adaptive, 
block time steps \citep{aarseth03}. 
Time steps of stars near the central  massive particle
(``black hole'') would become prohibitively small in such
a scheme. 
Therefore, the motion of stars near the black hole 
was treated via the chain regularization algorithm of
Mikkola and Aarseth \citep{mikkolaa90,mikkolaa93}.
This scheme selects an ordering for the $n$ particles in the chain 
and then regularizes the $n-1$ pairwise interactions
according to the standard Kustaanheimo-Stiefel (KS) transformation 
\citep{kustaans65},
such that motion in Cartesian space with a singular potential is 
transformed into harmonic motion in a 4D vector space.

Our chain method differs from the standard one
(which is incorporated in NBODY5 and NBODY6) in two ways. 
First, the massive body was always positioned at the bottom
of the chain; and second, the chain was ordered according 
to force, not distance.
The chain subsystem typically consisted of the black hole plus 
a small number of stars, $n-1\approx 10-20$.
The chain's center of mass was a pseudo-particle as seen by the
$N$-body code, and was advanced by the Hermite scheme in the same
way as an ordinary particle.
However the  structure of the chain (i.e. the forces from individual 
chain members) was resolved as necessary for the accurate integration
of nearby particles.
Furthermore the equations of motion of particles within the chain
included forces exerted by a list of external perturbers. 
Whether a given star was listed as a perturber was determined 
by a tidal criterion, with typically $\lap 10^2$ particles
satisfying the criterion.
Particles were denoted perturbers if their distance $r$
to the chain center of mass satisfied 
$r < R_{\rm crit1}= (m/m_{\rm ch})^{1/3} \gamma_{\rm min}^{-1/3} 
R_{\rm ch}$, where $m_{\rm ch}$ and $R_{\rm ch}$ represent the mass
and spatial size respectively of the chain, $m$ is the mass of
one star, and
$\gamma_{min}$ was chosen to  be $10^{-6}$; thus
$R_{\rm crit1} \approx 10^2 (m/m_{\rm ch})^{1/3}  R_{\rm ch} $. 
The chain was resolved for stars within a distance
$R_{\rm crit2}= \lambda_p R_{ch}$ from the chain's center of mass, 
where $\lambda_p=100$. 
In the standard chain implementations in 
NBODY5 and NBODY6 (with particles of nearly equal mass), 
$R_{\rm crit1} \approx R_{\rm crit2}$, 
while in our application these two radii are significantly different.

The Mikkola-Aarseth chain algorithm becomes ill-behaved in the
case of extreme mass ratios between the particles,
$\mh/m\gap 10^6$ (S. Mikkola, private communication).
\cite{aarseth03} advocates a time-transformed leap frog (TTL) 
to  treat such extreme mass ratios. 
The largest mass ratio in our simulations is $\sim 10^3$
for which the standard chain algorithm works perfectly well. 
Using a minimum tolerance of $10^{-12}$ for the chain's 
Bulirsch-Stoer integrator allowed us to reach typical relative
accuracies of $10^{-9}$. 

\section{Initial Conditions}

A crucial element of our method is the use of 
initial conditions that represent a precise steady state
of the collisionless Boltzmann equation.
Embedding or growing a massive particle in a pre-existing
stellar system can easily result in the formation of a density cusp
with slope similar to that predicted by Bahcall and Wolf (1976),
but for reasons having nothing to do with collisional relaxation.
For instance, as discussed above, adiabatic growth of a black hole produces
a density profile within the hole's sphere of influence
of $\rho\sim r^{-\gamma_{sp}}$,  $1.5\lap\gamma_{sp}\lap 2.5$; 
hence a cusp that
forms collisionlessly can easily mimic a $\rho\propto r^{-7/4}$
collisional cusp.
To avoid any possibility of non-collisional cusp formation
in our simulations, which would seriously compromise the 
interpretation,
we generated initial coordinates and velocities
from the steady-state phase space density $f(E)$ that
reproduces the Dehnen (1993) density law in the gravitational
potential including both the stars and the black hole.
The Dehnen model density satisfies $\rho(r) \propto r^{-\gamma}$
at small radii, and
the $f(E)$ that reproduces Dehnen's $\rho(r)$ in the presence
of a central point mass is non-negative for all $\gamma\ge 0.5$;
hence $\gamma=0.5$ is the flattest central profile that can be adopted
if the initial conditions are to represent a precise steady state.
As shown in Table 1, most of our runs used initial conditions
with this  minimum value of $\gamma$.
Henceforth we adopt units such that the gravitational constant $G$,
the total  stellar mass $M$, and the Dehnen scale  length $a$
are equal to one.

\begin{table}
 \begin{tabular}{c||c|c|c|c|c|c|c|c}
   &  $\gamma$ &  $N$  &  $M_{BH}/m$  &  $M_{BH}/M$  & $r_h$ &
 $\ln\Lambda$ & $T_r$  & $T_{max}$
 ($T_{max}/T_r$)  \\
 \hline  \hline
 1   &   1/2  &  250K  &  2500   &  0.01   &  0.26  & 8.0  &  1690  &
 1500 (0.887)   \\
 2   &   1/2  &  100K  &   500   &  0.005  &  0.19  & 6.6  &   454  &
 500  (1.101)  \\
 3   &   1/2  &  100K  &  1000   &  0.01   &  0.26  & 7.1  &   764  &
 800   (1.047)   \\
 4   &     1  &  100K  &  1000   &  0.01   &  0.17  & 6.6  &   227  &
 200  (0.881)  \\
 5   &   1/2  &  150K  &  1500   &  0.01   &  0.26  & 7.5  &  1034  &
 860 (0.832)  \\
 \end{tabular}
 \end{table}

Table 1 also gives the other important parameters of the 
$N$-body integrations.
The influence radius of the black hole $r_h$
was defined as the radius at which the enclosed stellar mass
at $t=0$ was equal to twice the black hole mass.
(This is equivalent to the more standard definition
$r_h=G\mh/\sigma^2$ when  $\rho\propto r^{-2}.$)
The relaxation time $T_r$ was computed at $t=0, r=r_h$ 
from the standard expression (eq. 2-62 of Spitzer 1987), setting
$\ln\Lambda = \ln(r_h\sigma^2/2Gm)$.
This definition of $\Lambda$ is equivalent to equating
$b_{max}$, the maximum impact parameter for encounters
in Chandrasekhar's theory, with $r_h$.
This choice is motivated by the expectation that $b_{max}$
for stars near the center of strongly inhomogenous stellar
system should be of order the radius at which the density falloff 
begins to abate, e.g. the core radius if there is a core
\citep{Maoz:93,Merritt:01}.
In our simulations, this radius  is of order the Dehnen scale length
$a\approx {\rm a\ few}\times r_h$ at $t=0$, decreasing 
to a fraction of $r_h$ after formation of  the collisional cusp; 
hence a choice of $b_{max}\approx r_h$
seems appropriate.
We stress that Chandrasekhar's $\ln\Lambda$ is a poorly-defined quantity
in strongly inhomogeneous and evolving systems, and our
choice is at best approximate. 
Nevertheless, we will see that the time scaling determined
by this choice of $\ln\Lambda$ results in very good correspondence
between the evolution rates seen in the $N$-body and Fokker-Planck
models.

\section{Fokker-Planck Models}

We compared the $N$-body evolution with the predictions
of the time-dependent, orbit-averaged, isotropic Fokker-Planck equation:
\begin{subequations}
\begin{eqnarray}
& & 4\pi^2p(E){\partial f\over\partial t} = -{\partial F_E\over\partial E}
 ,\\
& & F_E(E,t) = -D_{EE}{\partial f\over\partial E} - D_Ef,\\
& & D_{EE}(E) = 64\pi^4G^2m^2\ln\Lambda\times  \nonumber \\ 
& & \left[q(E) \int_{-\infty}^E dE'f(E') + 
\int_E^\infty dE' q(E')f(E')\right], \nonumber \\
& & D_E(E) = -64\pi^4G^2m^2\ln\Lambda\int_E^\infty dE'p(E')f(E')
\end{eqnarray}
\label{eq:fp}
\end{subequations}
\noindent
(e.g. Cohn 1980; Spitzer 1987).
Here $p(E) = 4\sqrt{2}\int_0^{r_{\rm max}(E)} r^2 \sqrt{E-\Phi(r)} dr
=-\partial q/\partial E$
is the phase space volume accessible per unit of energy,
the upper integration limit is determined via $-\Phi(r_{\rm max}) = E$,
and $E=-v^2/2 - \Phi(r)\ge 0$ is the binding energy per unit mass
(hereinafter the ``energy.'')
Since the gravitational potential changes as the stellar distribution
evolves, equation (\ref{eq:fp}) should also contain a term
describing the adjustment in $f$ due to 
changes in $E$ as the gravitational potential evolves
 (e.g. Spitzer 1987, eq. 2-86).
However changes in $\rho(r)$ only take place well within
$r_h$ in our simulations, and the potential is dominated by the (fixed) mass
of the black hole in this region.
Hence, assuming a fixed potential in the Fokker-Planck equation
is an excellent approximation for our purposes.

When scaling the Fokker-Planck  results  to the $N$-body results,
the only free parameter is $\ln\Lambda$.
We used the values  given in Table  1.

\section{Results}

As the cusp develops, the density of stars at $r\lap r_h$
increases.
This is illustrated in Figure 1, which shows the mass in stars
within a radial distance $0.1 r_h$ from the black hole
as a function of time for each of the runs.
(Distances were defined with respect to the instantaneous
position of the black hole particle; the latter wanders like
a Brownian particle, but the density peak tends to remain centered on the
black  hole as it moves.)
For comparison, we also show in Figure 1
the  same quantity as computed from the
Fokker-Planck equation.
The time scaling of the Fokker-Planck equation was set
using the value of $\ln\Lambda$ given in Table 1 -- no
adjustments were made to optimize the fit.
(We note that integrations like  these  could in principled 
be used to {\it evaluate} $\ln\Lambda$.)
While there are hints of systematic differences in some
of  the runs, overall the correspondence is very good: 
clearly, the $N$-body evolution is quite close to what 
is predicted from the Fokker-Planck equation.

\begin{figure}
\plotone{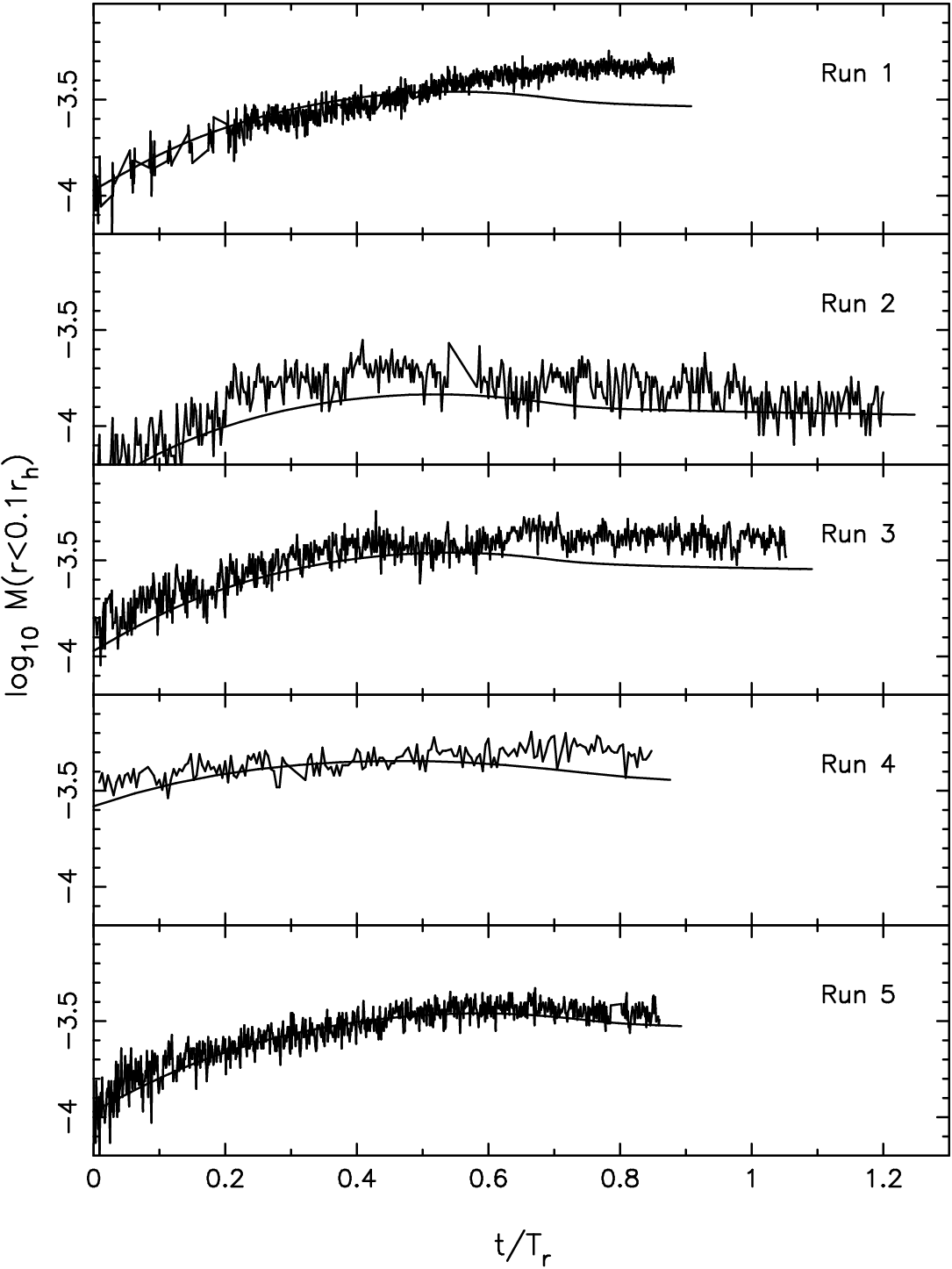}
\caption{Evolution of the mass in stars within a distance
$0.1 r_h$ from the black hole, where $r_h$ is the influence
radius of the black hole measured at time zero.
Noisy curves are from the $N$-body runs; smooth
curves are solutions to the Fokker-Planck equation.
  \label{fig1}}
\end{figure}

The $N$-body runs in Figure 1 exhibit a range in $N$ from
$0.25\times 10^6$ (Run 1) to $0.1\times 10^6$ (Runs 2-4), showing
that the correspondence between $N$-body and Fokker-Planck results
remains good over at least a modest range in particle number.
Figure 1 also suggests that each of the integrations has reached
an approximate steady state with regard to changes in the density
at the final time step.

Having demonstrated the reliability of the time scaling of the
Fokker-Planck equation,
we can make a more detailed comparison with the $N$-body models.
Figure 2 shows the evolution of the stellar density
$\rho(r)$ in Run 1 compared with the Fokker-Planck prediction.
The $N$-body density was computed from snapshots of the particle positions
(no time averaging) using MAPEL,
a maximum penalized likelihood algorithm
\citep{Merritt:94,MT:94}.
The radial coordinate was defined again as distance from 
the black hole.
The correspondence between $N$-body and Fokker-Planck results
is again quite good; the  only systematic difference appears at
very small radii ($r\lap 0.01 r_h$) where the particle numbers 
are too small for reliable estimates of $\rho$.
The final cusp is well represented by $\rho\propto r^{-1.75}$
at $r\lap 0.1 r_h$,
both in Run 1 and in the other integrations (not illustrated here).

\begin{figure}
\plotone{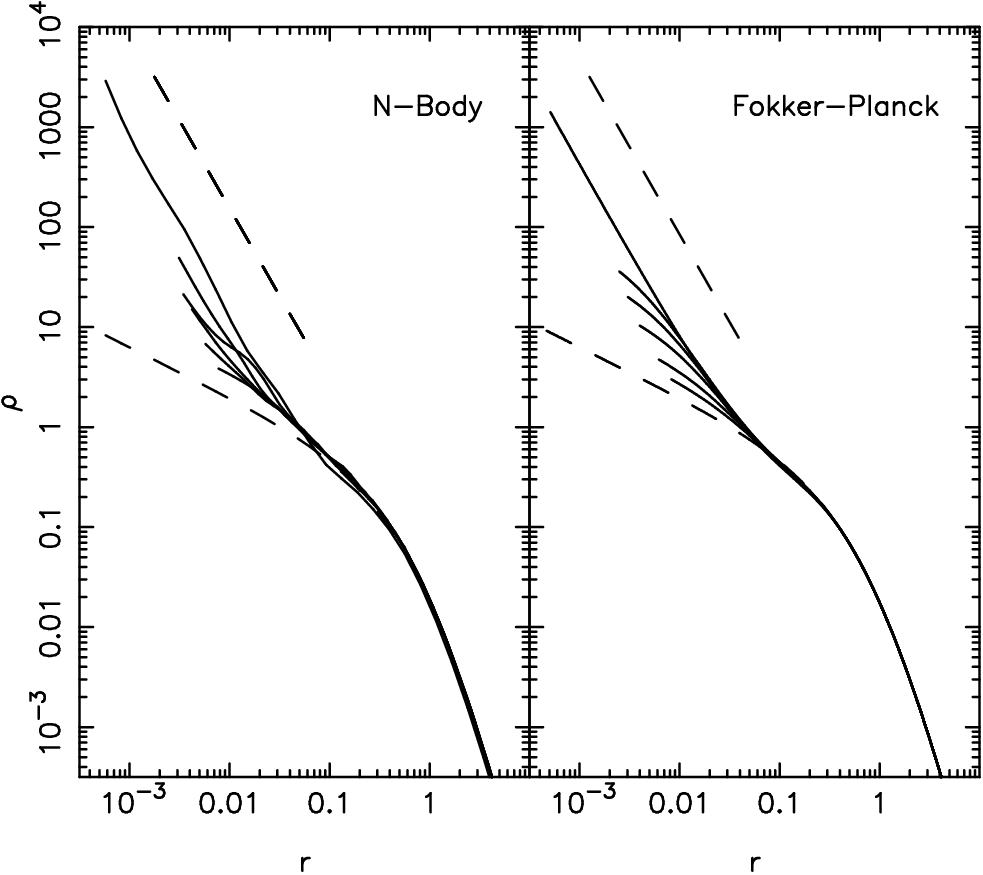}
\caption{Evolution of the mass density profile around the black
hole.
Left panel: $N$-body; $\rho(r)$ was estimated from the particle
positions at times $t=100, 200, 300, 400, 500, 1500$ via
a maximum penalized likelihood algorithm.
Right panel: Densities predicted from the Fokker-Planck equation
at the same times; scaling of the time unit used the value
of $\ln\Lambda$ given in Table 1.
Lower dashed curves show the density at $t=0$; upper
dashed curves show $\rho\propto r^{-7/4}$, the
asymptotic solution to the Fokker-Planck equation.
  \label{fig2}}
\end{figure}

The large number of particles in our $N$-body experiments permits
us to go one step deeper in comparing $N$-body with 
Fokker-Planck results.
It is possible to  extract estimates of $f(E)$ from the $N$-body
data sets.
We  did this as follows.
Snapshots of the particle positions and velocities were stored
at each $N$-body time unit; such data sets are essentially
uncorrelated at radii near $r_h$.
Roughly $70$ of these snapshots were then combined into a
single data file, giving an effective $N$ of $\sim 10^7$.
From the combined data set we computed an estimate of
the gravitational potential using standard expressions,
then computed the phase-space volume element $p(E)$ defined
above.
The particle energies were also computed, and a histogram
constructed of $N(E)$.
Finally, we used the relation $f(E)=N(E)/4\pi^2p(E)$ to compute 
an estimate of the phase-space density, assuming an isotropic distribution
of particle velocities.

\begin{figure}
\plotone{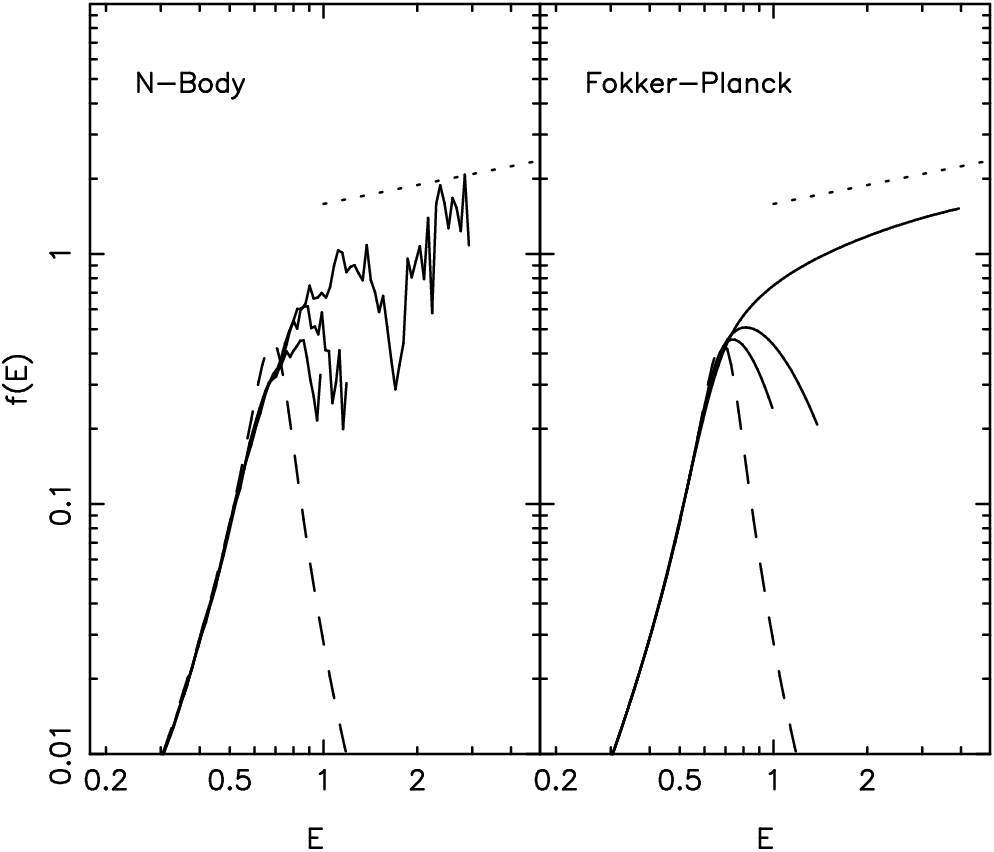}
\caption{Evolution of the phase-space density of 
stars  around the black hole in Run 5.
Left panel: $N$-body; $f(E)$ was estimated from the particle
energies at times $t=180, 300, 600$
as described in the text.
Right panel: Densities predicted from the Fokker-Planck equation
at the same times; scaling of the time unit used the value
of $\ln\Lambda$ given in Table 1.
Lower dashed curves show $f(E)$ at $t=0$; upper
dashed curves show $f\propto E^{1/4}$,
the asymptotic solution to the Fokker-Planck equation.
  \label{fig3}}
\end{figure}

Figure 3 shows the results for Run 5 ($N=1.5\times 10^5$, $\gamma=0.5$).
While $f(E)$ is clearly harder to estimate than $\rho(r)$ --
it is effectively a $3/2$ derivative of $\rho$ and hence
noisy -- we can
see from Figure 3 that the $f$ extracted from  the  $N$-body
runs evolves in a very similar way to the $f$ computed via
the Fokker-Planck equation, and that its
steady-state  form is consistent (modulo the noise) with
the Bahcall-Wolf solution $f\propto E^{1/4}$ at late times.

\section{Discussion}

Our results contribute to the growing body of
literature demonstrating the feasibility of direct-force
$N$-body techniques for simulating the evolution of dense
stellar systems on collisional time scales.
Two landmark studies in this vein  were the recovery
of gravothermal oscillations using the GRAPE-4 special
purpose computer by Makino (1996),
and the computation of the parameters of
core collapse by Baumgardt  et al. (2003) using the GRAPE-6.
Our  study extends  this work to  the yet more difficult
case of star clusters around massive singularities,
in  which accurate  and efficient computation of the
forces is particularly critical.
We have shown that particle numbers accessible to
a single special-purpose computer ($\lap 10^6$),
combined with a regularization scheme for close
encounters to the black hole,
are sufficient to test predictions derived from
a macroscopic formulation of the evolution equations,
and that  the two approaches indeed make consistent predictions
about  the formation of a cusp on relaxation time scales.

This success points the way to even more challenging
$N$-body studies of galactic nuclei.
For instance, the ``loss cone'' problem -- the rate
at which a central black hole captures or destroys
stars -- has  so far been the almost exclusive
domain of Fokker-Planck treatments.
Verifying the Fokker-Planck predictions via
direct $N$-body simulations is feasible but will
require particle numbers large enough that
loss-cone refilling takes place on a time scale
long compared with orbital periods.
This implies $N>10^6$ \citep{milosm03},
currently achievable via distributed hardware
\citep{Dorband:03}.
Related problems requiring comparably large $N$
are the long-term evolution of binary supermassive
black holes, 
and the  Brownian motion of black holes in
dense nuclei.

\bigskip\bigskip

We are indebted to Sverre Aarseth and Seppo Mikkola who
were tireless in their advice and support during the coding
of the chain regularization algorithm.
We  gratefully acknowledge useful  discussions with
Marc Freitag and Simon Portegies Zwart.
This work was supported via grants NSF AST 02-0631, 
NASA NAG5-9046 and HST-AR-09519.01-A
to DM, and by the German Science Foundation
(DFG) via grant SFB439 at the University of Heidelberg to RS and MP.
MP also acknowledges financial support from the Funda\c{c}ão para a Ciência 
e Tecnologia (FCT), Portugal, through grant SFRH/BD/3444/2000.

\end{document}